\documentclass[aps,prapplied,reprint,a4paper,superscriptaddress,amsmath,amsfonts,amssymb, longbibliography,floatfix, 
]{revtex4-2}
\usepackage{bm}
\usepackage[caption=false]{subfig}
\usepackage{tikz}
\usetikzlibrary{shapes.misc,positioning,shapes}
\usetikzlibrary{patterns}
\usepackage{array, multirow}
\usepackage[colorlinks,linkcolor=blue,citecolor=blue]{hyperref}%
\usepackage{relsize}

\usepackage{times}

\usepackage[normalem]{ulem}

\usepackage[english]{babel}
\usepackage{blindtext}

\newcolumntype{P}[1]{>{\centering\arraybackslash}p{#1}}

\begin{document}
\title{Designing fast quantum gates using optimal control with a reinforcement-learning ansatz}

\author{Bijita Sarma}
\email{bijita.sarma@fau.de}
\affiliation{Department of Physics, Friedrich-Alexander-Universit\"at Erlangen-N\"urnberg, 91058 Erlangen, Germany}
\author{Michael J. Hartmann}
\affiliation{Department of Physics, Friedrich-Alexander-Universit\"at Erlangen-N\"urnberg, 91058 Erlangen, Germany}
\affiliation{Max Planck Institute for the Science of Light, 91058 Erlangen, Germany}


\begin{abstract}
Fast quantum gates are crucial not only for the contemporary era of noisy intermediate-scale quantum devices but also for the prospective development of practical fault-tolerant quantum computing. Leakage errors, which arise from data qubits jumping beyond the confines of the computational subspace, are the main challenges in realizing non-adiabatically driven, fast gates. In this work, we propose and illustrate the usefulness of reinforcement learning (RL) to generate fast two-qubit gates in practical multilevel superconducting qubits. In particular, we show that the RL controller offers great effectiveness in finding piecewise constant gate pulse sequences that act on two transmon data qubits coupled by a tunable coupler to generate a controlled-Z (CZ) gate with a gate time of 10 ns and an error rate of $\sim 4\times 10^{-3}$. Using a gradient-based method to solve the same optimization problem often does not achieve high fidelity for such fast gates. However, we show that using the gate pulses discovered by RL as an ansatz for the gradient-based controller can substantially enhance fidelity compared to using RL alone. While for a 10 ns pulse, this improvement is marginal, the combined RL + gradient approach decreases the gate errors below $10^{-4}$ for a gate of length 20 ns.
\end{abstract}

\maketitle
\section{Introduction}
As we are inching closer to building practical quantum computers, the need for developing fast quantum gates has become increasingly important~\cite{Nielsen2010Dec}. This is critical in the present noisy intermediate-scale quantum (NISQ) era, allowing the reliable execution of quantum algorithms despite the intrinsic noise and fragility of qubits and facilitating fault tolerance through the effective implementation of error-correcting gates in large quantum systems~\cite{Cerezo2021SepVQE, Preskill2018Aug, Arute2019Oct, Kjaergaard2020Mar, Rosenblum2018Jul, DiCarlo2009Jul, Andersen2020Aug}. Fault tolerance implies that, as long as the error rates of the physical qubits remain below a certain threshold, quantum computing systems, together with quantum error correction (QEC) and logical gate operations, can be utilized for useful computation~\cite{Ma2020Aug, Krinner2022May, Knill1997Feb, Grimsmo2021Jun}. One of the crucial requirements for efficient QEC is the realization of fast and efficient quantum gates~\cite{google2023suppressing}. 

As a hardware platform for quantum computing, superconducting circuits have shown remarkable developments and are considered promising for the construction of large-scale quantum devices. However, designing fast quantum gates with high fidelity remains a major challenge. Since superconducting qubits are in fact multilevel systems, the most significant challenge in achieving fast high-fidelity two-qubit gates is avoiding leakage outside the computational subspace during gate execution~\cite{Chen2016Jan,Bultink2020Mar, Fazio1999Dec, Chou2023Jul}. These leakage errors are extremely difficult to minimize, and methods to prevent them restrict the amplitude of control pulses, consequently extending the gate duration.  Another major issue for the accurate functioning of large-scale superconducting systems is qubit crosstalk caused by residual $ZZ$ interactions, leading to undesired disruptions in two-qubit gate operations. A method of minimizing crosstalk involves enhancing the hardware architecture, for instance, by employing qubits with different anharmonicities to generate a crosstalk cancelation effect through quantum interference~\cite{Yan2018Nov,Heunisch2023Jun,Mundada2019Nov}. The more experimentally practical approach is to position the qubits in a highly dispersive regime with significant detunings. However, in these configurations, transitioning from a state of large detuning to an operating frequency zone with lesser detuning results in slower gate operations. In this work, we present a technique to implement a rapid two-qubit gate by starting from a condition with negligible residual coupling. Despite steep ramps, the optimization process, which leverages reinforcement learning (RL) combined with optimal control, facilitates the realization of an accelerated two-qubit $CZ$ gate.

When dealing with global optimization problems of complex, non-convex, and non-linear systems, machine learning (ML) in combination with deep learning has recently been shown to be extremely successful and is considered highly versatile for a wide range of tasks~\cite{Goodfellow-et-al-2016, silver_mastering_2016, silver_mastering_2017}. 
RL is a type of ML that is particularly suited for learning to control sequential decision-making problems~\cite{sutton2018reinforcement,Krotov1995Oct,Caneva2011Aug}. In the last couple of years, RL has been utilized to find control protocols for some interesting quantum problems~\cite{Krenn2023Jan, Gebhart2023Mar}. It was first demonstrated for the optimization of quantum phases~\cite{Bukov2018Sep} and QEC~\cite{Fosel2018Sep}, and more recently we have seen its applications in other areas, in particular, in quantum state engineering~\cite{Porotti2019Jun}, quantum pulse and gate design~\cite{Sarma2022Nov, Ding2023Sep}, quantum feedback control~\cite{Borah2021Nov, Wang2020Sep, Borah2023Nov} etc. RL controls have also been used in real laboratory experiments recently with quantum systems, demonstrating their potential for challenging decisions and their adaptability to control such systems in real time~\cite{Sivak2023Apr, Reuer2023Nov}.

\begin{figure}[t]
    \centering
    \includegraphics[width=1.0\linewidth]{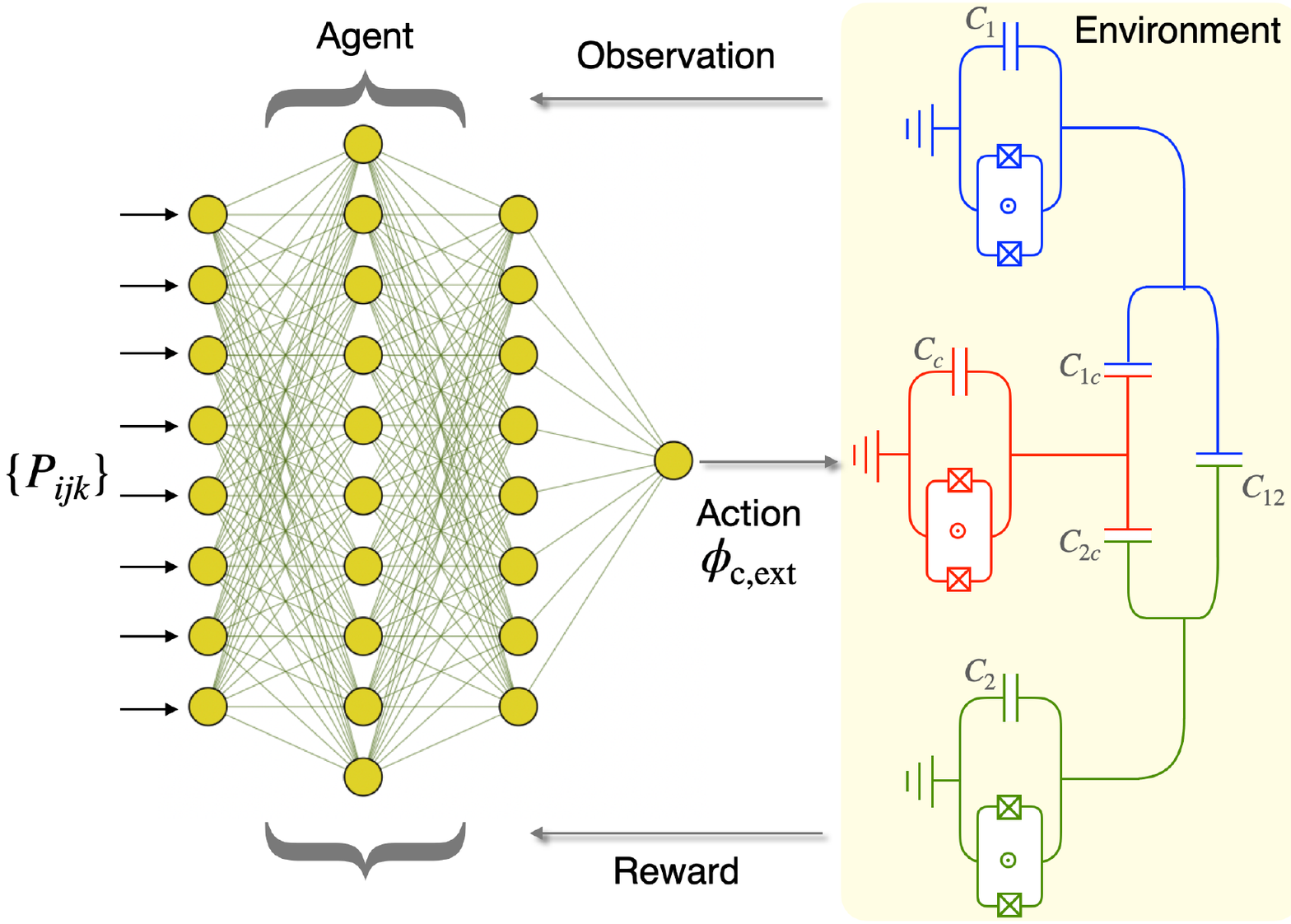}
    \caption{Schematic diagram showing the RL-controlled gate implementation in a three-qubit tunable coupler circuit. The circuit (right), which constitutes the RL-environment, is made of three transmon superconducting qubits, each modeled as a multi-level anharmonic oscillator, with capacitive nearest-neighbor and next-nearest-neighbor couplings. The RL agent is fed with observations $\{P_{ijk}\}$, i.e.~the computational and leakage-space populations. Based on these observations, the agent exerts actions to control the coupler frequency $\omega_c(t)$, and receives a reward or penalty in terms of the gate infidelity $\mathcal{I} = 1 - \mathcal{F}$, where $\mathcal{F}$ is the gate fidelity at the end of the sequence. 
    }
    \label{fig:fig1}
\end{figure}

Quantum gate preparation tasks have traditionally been addressed using various optimal control methods, such as gradient ascent pulse engineering (GRAPE), which, as its name implies, employs gradient-based optimization to minimize a loss function, typically the gate infidelity~\cite{Khaneja2005Feb, Jager2014Sep}, or using Krotov's method by iteratively updating control fields~\cite{Krotov1993}. Such methods often depend on the system dynamics being differentiable, thereby necessitating an accurate understanding of the quantum system model. In contrast, RL algorithms are commonly employed as a model-free method that only needs the output data as the observation to be used to map into control sequences. Furthermore, it is highly adaptable to variations in system parameters, which helps to mitigate model bias. This adaptability is particularly useful if the RL agent is initially trained on a simulator and adjusted based on new insights obtained from experimental data to refine the control sequences. With recent advances in high-speed electronic components such as Field Programmable Gate Array (FPGA), RL can now be fully applied to experiments using real-time data~\cite{Sivak2023Apr, Reuer2023Nov}. While gradient-based methods are typically more straightforward than RL and can achieve the desired accuracy with considerably less effort, they heavily rely on the initial parameter values of the gradient-based optimizers, especially for complex tasks like quantum gate preparation discussed in this work. We show that by combining these two approaches, viz.~RL and optimal control, it is possible to take advantage of the strengths of both by first optimizing pulses with RL and then refining them further using optimal control techniques.

\section{Model and Methods}
We consider a tunable coupler superconducting circuit setup, (see Fig.~\ref{fig:fig1}), to design an ultrafast two-qubit controlled-Z (CZ) gate.
Besides being an entangling gate that can be used to generate a universal gate set, the CZ gate is a core operation in QEC with surface codes~\cite{Krinner2022May,google2023suppressing}. For engineering high-performance, large-scale quantum processors, tunable superconducting circuits have gained prominence, particularly due to their recently recognized capability for on-demand on-off switching of couplings between qubit pairs via frequency modulation through external fluxes~\cite{google2023suppressing,Sung2021Jun, Stehlik2021Aug,Heunisch2023Jun,Ding2023Sep}. 
This flexibility allows for precise control over the interactions within the system, making it a valuable resource
for implementing high-fidelity gates.

The RL-based optimization protocol is depicted in Fig.~\ref{fig:fig1}, where the problem of two-qubit gate design with a tunable coupler superconducting framework is embedded in the RL workflow. The RL agent (shown on the left) is essentially an artificial neural network model that is responsible for deciding the control sequences (called the actions, $\vec{a}$) by optimizing the weights, $\vec{\theta}$ of the model. These optimizations are directed through the scalar signal of rewards, $\mathcal{R}$ received by the RL agent from the RL-environment given it observes some partial information of the system after the application of the control at every step of iteration. These are called the observations, $\vec{s}$ of the RL, and the set of rules it learns by optimizing the parameters $\vec{\theta}$ is called the policy, $\pi(\vec{a}|\vec{s})$ of the RL agent, where $\pi(\vec{a}|\vec{s})$ represents a conditional probability distribution of $\vec{a}$ given $\vec{s}$. In this case, the RL-environment is formed by the circuit shown on the right of Fig.~\ref{fig:fig1}, comprising two data qubits and a coupler qubit, all of which are modeled as transmon qubits. Explicitly, the observation of the RL agent consists of the computational as well as the leakage-state populations in the eigenstates of the three qubits at the idle points of the gate, i.e., $\vec{s} = \{P_{ijk} \}$, where $\{i, j, k\}$ denotes the qubit 1, coupler and qubit 2 respectively. The actions of the RL agent are choices of the tunable coupler frequency, $\omega_c$,  which are realized by the external flux, $\phi_{c, \text{ext}}$ applied to the coupler, therefore $\vec{a} =\{\omega_c\} \leftarrow \phi_{c, \text{ext}}$. The reward, $\mathcal{R}$ is considered as a function of the process infidelity of the CZ gate defined by $\mathcal{R} = -\log_{10}(1 - \mathcal{F})$, where $\mathcal{F}$ is the gate fidelity at the end of the sequence.

The learning process can be divided into iterations called episodes, each with a total duration of $\tau$, which is further segmented into sequences with a duration of $t' = \tau/n$, where $n$ is the number of control steps in the episode. The episode time $\tau$ is equivalent to the gate time in our case.
If we consider a control problem with $N_{a}$ control parameters over $n$ control steps in each episode, the complexity of the problem scales exponentially with $n$ as $ \prod_{i=1}^{N_a} n_{i}^{n}$, where $n_i$ is the number of choices for the $i$-th control parameter, considering a discrete control problem. For the problem under study $N_a = 1$, corresponding to the control parameter $\omega_c$, for which the complexity of the problem scales as $n_{1}^n$, where $n_{1}$ is the number of choices of the control parameter $\omega_c$. 
Instead of discrete controls, we consider continuous approximations to stepwise constant shapes of $\omega_c$. This choice of control pulses is motivated by the fact that they are typical pulses generated by arbitrary waveform generators in current experimental setups. 

Despite the fact that we have a single control parameter for the RL agent to learn, this problem turned out to be a formidable task for the RL to learn and we have found that a sophisticated RL algorithm developed in the last few years needs to be employed. Effectively, we used the recently proposed Soft-Actor-Critic (SAC) algorithm for optimization of the RL policy~\cite{sac_paper}. The SAC algorithm is an actor-critic RL algorithm based on the concept of entropy regularization. The policy $\pi$ is trained to maximize a trade-off between expected return and entropy. This trade-off determines the balance between exploration and exploitation. The algorithm provides a bonus reward at each time step proportional to the entropy of the policy.
This makes the RL-policy to spawn actions as randomly as possible due to the inherent stochasticity of the policy, encouraging the agent towards more exploration, prevention of premature convergence to sub-optimal solutions, and accelerated learning. The optimal policy $\pi^*$ is defined as the policy that maximizes the expected return while also maximizing entropy, given by
\begin{align}
\nonumber
\pi^{*} \ = \  \mathrm{arg} \max_\pi\underset{\tau \sim \pi}{\mathbb{E}} & \sum_{t=0}^{\infty} \gamma^t [\mathcal{R}(s_t, a_t, s_{t+1})\\
 +& \alpha \mathcal{H}\left(\pi(\cdot |s_t)\right)],    
\end{align}
where $\mathbb{E}_{\tau \sim \pi}$ denotes the expectation value over trajectories $\tau$ generated by following the policy $\pi$. $\gamma^t$ is the discount factor raised to the power of $t$, where $\gamma$ is a parameter between 0 and 1, representing how much the agent values future rewards relative to immediate rewards. $\mathcal{R}(s_t, a_t, s_{t+1})$ represents the immediate reward obtained when taking action $a_t$ in state $s_t$ and transitioning to state $s_{t+1}$. The $\alpha \mathcal{H}\left(\pi(\cdot|s_t)\right)$ term involves the entropy $\mathcal H$ of the policy $\pi$ at state $s_t$, weighted by a hyper-parameter $\alpha$, that regulates stochasticity of the policy, and encourages randomness in the actions taken by the RL agent [see Appendix A for details].

The tunable coupler circuit depicted in Fig.~\ref{fig:fig1} is described by the Hamiltonian (considering $\hbar=1$ hereinafter),
\begin{eqnarray}
\label{eqn:hamiltonian}
\nonumber
    H &=& \sum_{i=1,c,2} \left(\omega_i b_i^\dagger b_i + \frac{\alpha_i}{2} b_i^\dagger b_i^\dagger b_i b_i\right) + g_{12} \bigl(b_1 + b_1^\dagger \bigr) \bigl(b_2 + b_2^\dagger \bigr)\\
    &+&  \sum_{i=1, 2} g_{ic} \bigl(b_i + b_i^\dagger\bigr) \bigl(b_c + b_c^\dagger\bigr) \ , 
\end{eqnarray}
where $b_i (b_i^\dagger)$ with $i=1,c,2$ describe the bosonic annihilation (creation) operators for the transmon qubits ($i=1,2$) and the coupler, where the qubits are considered as weakly anharmonic oscillators possessing multiple energy levels, with anharmonicities given by $\alpha_i$. The qubits interact with one another through capacitive coupling,
where the next-nearest-neighbour coupling capacitance, $C_{12}$, is smaller than the nearest-neighbour coupling capacitances, $\{C_{1c},\ C_{2c}\}$, which in turn are small compared to the transmon qubit capacitances $\{C_1,\ C_c,\ C_2\}$. This leads to the fact that the circuit analysis can be treated perturbatively. 
The circuit is initialized at the idle point with the eigenstates $|ijk\rangle$ (where $\{i,\ j,\ k\}$ label the qubit $1$, coupler and qubit $2$ respectively) before application of the gate and is returned back to after the completion of the gate. At the idle point, these instantaneous eigenstates have maximum overlap with the bare states of the circuit, with a slight hybridization between the data qubits due to residual coupling~\cite{Heunisch2023Jun}. The experimentally relevant computational subspace for the two-qubit gates between the data qubits consists of the eigenstates at the idle points $|ik\rangle$ $=$ $|00\rangle$, $|01\rangle$, $|10\rangle$ and $|11\rangle$, where the coupler is considered to be always in the ground state. All other eigenstates constitute the leakage subspace.  

\begin{figure}[t]
    \centering
    \includegraphics[width=1.0\linewidth]{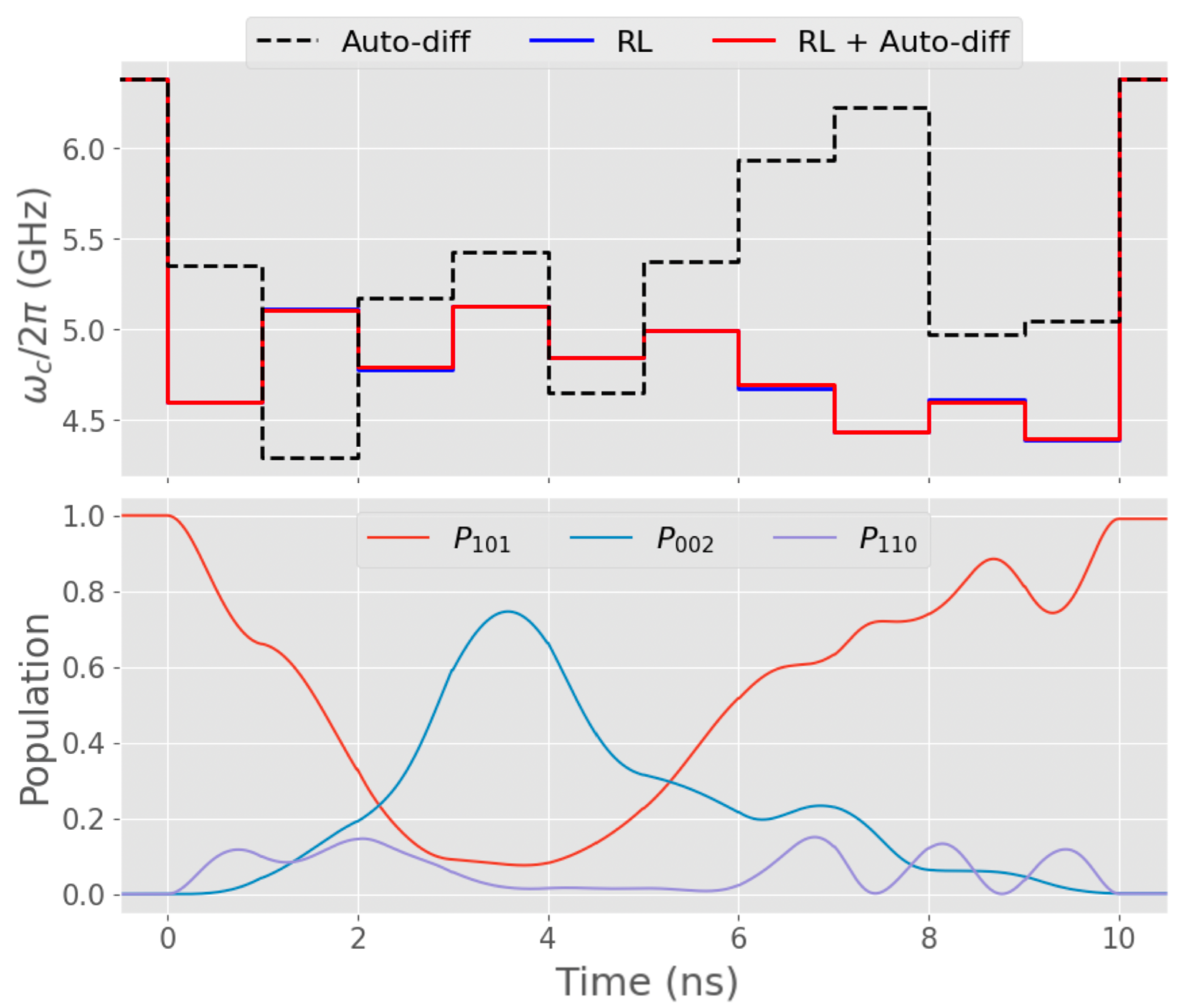}
    \caption{(Top panel) Piecewise constant gate control sequence for the coupler qubit frequency obtained from RL (blue), automatic differentiation (in black dashed), and combined RL and Auto-differentiation (red). Before implementation of the gate, the qubits are started at the idle point with negligible crosstalk, and also brought to the same resting configuration after the gate. (Bottom panel) The corresponding population of the significantly occupied computational and leakage-space states, $P_{ijk}$, are shown for the RL + Auto-diff-derived pulses throughout the gate evolution [See the main text for further detail].}
    \label{fig:fig-2}
\end{figure}

We aim to design the CZ gate utilizing the transverse qubit-qubit coupling to induce a phase of $e^{i\pi}$ in the computational state $|101\rangle$ by using nonadiabatic transitions to the non-computational eigenstate $|002\rangle$ and back. At the idle point, we consider the qubits to be in the highly dispersive regime where the detuning between the coupler and the qubits is large compared to their mutual couplings, so that both the transverse and longitudinal couplings between the qubits are negligible. 
We bias the qubits  and coupler at the frequencies of $\omega_1/2\pi = 4.2\,\mathrm{GHz}$, $\omega_2/2\pi = 5.2\,\mathrm{GHz}$ and $\omega_c/2\pi = 6.38\,\mathrm{GHz}$. This results in negligible transverse and longitudinal ZZ couplings [see Appendix B for details]. 
The other parameters are, $\alpha_1/2\pi = -200\,\mathrm{MHz}$, $\alpha_c/2\pi = -100\,\mathrm{MHz}$, $\alpha_2/2\pi = -200\,\mathrm{MHz}$, $g_{1c}/2\pi = 85\,\mathrm{MHz}$, $g_{2c}/2\pi = 85\,\mathrm{MHz}$, and $g_{12}/2\pi = 7\,\mathrm{MHz}$.
 
Starting at this dispersive coupling limit, a CZ gate can be obtained by first tuning the frequency of qubit 1 to $\omega_1 = \omega_2 + \alpha_2$, so that the levels $|101\rangle$ and $|002\rangle$ become resonant, and then tuning the coupler frequency close to the data qubit frequencies. Holding the coupler frequency at this point for the time of one oscillation between these two states, the target unitary $U_{\text{CZ}} = \text{diag}(1, 1, 1, -1)$ can be achieved up to single-qubit phases, which can be virtually compensated for. 
However, the gate time for such a Rabi-oscillation-based operation is long as it is given by $t_{\mathrm {gate}} = \pi/\zeta_{XX}$, where $\zeta_{XX}$ is the transverse coupling rate, as it needs to satisfy the adiabaticity condition $\int {\zeta_{XX}(t) dt} \gg 1$~\cite{Genov2023Feb}. As discussed previously, the RL agent's task is to tune the coupler frequency $\omega_c (t)$ throughout the duration of the gate within the given interval, $\omega_c(t)/2\pi = [4.2, 6.38]~ \text{GHz}$ to maximize the fidelity of the gate at the end of the gate. Given the extensive exploration abilities of neural networks, RL is expected to outperform standard gradient-based methods that struggle with shallow minima in intricate control tasks. Nevertheless, for challenging problems like the one discussed here, RL might also perform suboptimally, oscillating between minima with nearly identical returns, and frequently abandoning such solutions in pursuit of further exploration, ultimately deviating from the original strategy. We show that using the optimum discovered by an RL stage as the initial ansatz for an optimal control gradient stage, allows us to find a better solution than the one discovered by RL in terms of achievable fidelity within short gate times.

\begin{figure}[t]
    \centering
    \includegraphics[width=0.9\linewidth]{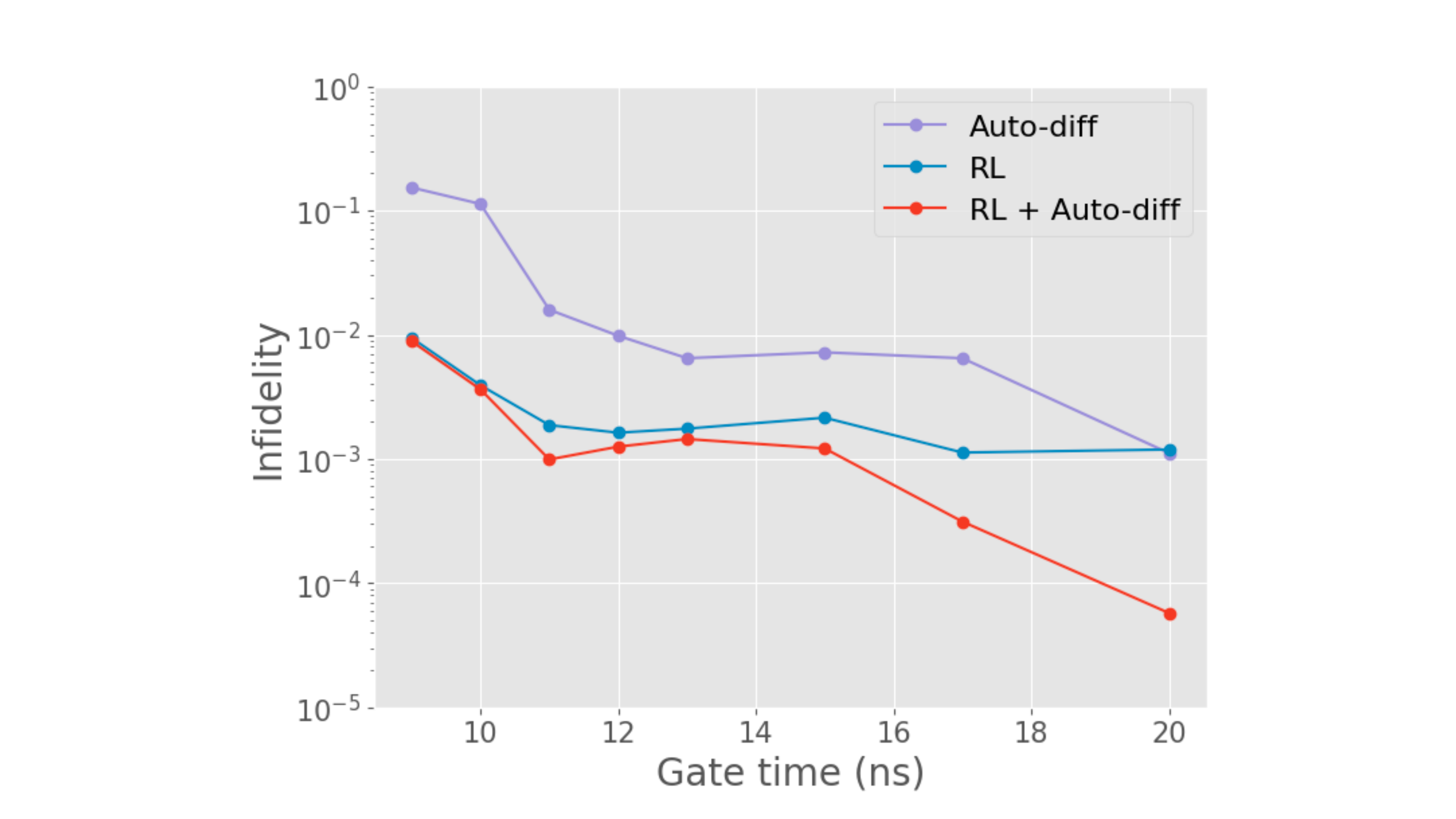}
    \caption{Comparison of the performance of the three optimization methods, viz.~Auto-diff (purple), RL (blue) and RL combined with Auto-diff (red), in relation to variations in gate time. }
    \label{fig:fig-3}
\end{figure}

\section{Results and Discussions}
The results are shown in Fig.~\ref{fig:fig-2}, which illustrates the findings related to the identification of a 10-ns-long  CZ gate. The upper panel of the figure presents the control pulses, with the results derived from the gradient with automatic differentiation (Auto-diff), RL, and combined RL + Auto-diff methods. The gradient descent technique seeks to fine-tune the values of $\omega_c(t)$ over the period $t=[0, 10]$ ns to reduce the gate's infidelity at the end, while RL utilizes an iterative optimization process, modifying actions based on the feedback obtained after each action $\omega_c[t]$ in order to increase the gate fidelity using a neural network policy developed through extensive trial-and-error learning. We find that the simple gradient-based method with an intuitive initial ansatz for $\omega_c(t)$ could not find a good solution for the 10 ns CZ gate with a fidelity of about 90\% (averaged over runs with different initial ansatz), while the one discovered by RL yields a fidelity of 99.60\%. The fidelity is marginally enhanced to 99.63\% by employing the RL-derived solution as an ansatz for the gradient-based method. This distinction becomes considerably more prominent for gates with extended gate durations, as depicted in Fig.~\ref{fig:fig-3}, which illustrates the infidelity as a function of gate times ranging from 10 to 20 ns discovered through the three approaches. RL + Auto-diff optimization shows an overall improved strategy resulting in consistently better fidelities reaching a value higher than $99.99\%$ for a 20 ns gate time. Although gate sequences shorter than this achieve high speed by partially populating the leakage channels at the expense of lower fidelity~\cite{Khani2009Dec}, the gate with a duration of the order of 20 ns falls below the surface code Pauli error threshold of $\sim0.01$, as well as leakage error threshold of $\sim10^{-4}$~\cite{Suchara2015Sep} [see Appendix F for an analysis of leakage population with respect to gate time].
Nonetheless, if the leakage is above threshold, there are several specific techniques, such as leakage reduction units or teleportation, that can be applied to get the leakage under control and bring the qubits back into the computational space before regular error correction cycles start~\cite{Aliferis2007Jan, Suchara2015Sep, Werninghaus2021Jan, McEwen2021Mar, Miao2023Dec, Hyyppa2024Sep, Chen2024Sep}. 
Also, since the decoherence time for state-of-the-art transmons is of the order of $\tau' \sim 60\,\mu\mathrm{s}$, this leads to error rates characterized by $\varepsilon_{{\tau}'} = 1 -\exp (-t_{\text{gate}}/\tau')\approx 3\times10^{-4}$ for $t_{\rm gate}=20$ ns~\cite{Sung2021Jun}. Therefore, executing such a rapid gate operation is feasible well before substantial information loss occurs due to decoherence.

The bottom panel of Fig.~\ref{fig:fig-2} shows the populations of the $|101\rangle$, $|002\rangle$ and $|110\rangle$ states during the application of the gate pulses found above. The RL agent finds optimal conditions for the pulse variation between the states $|101\rangle$ and $|002\rangle$ to acquire the desired phase. Throughout the gate operation, the leakage into the coupler is greatly minimized due to the alternating positive and negative shifts in qubit frequency, with a further reduction observed by the end of the gate. The extent of leakage is also reduced substantially for longer-duration gates, as shown in Appendix C.

\begin{figure}[!hbt]
    \centering
    \includegraphics[width=1.0\linewidth]{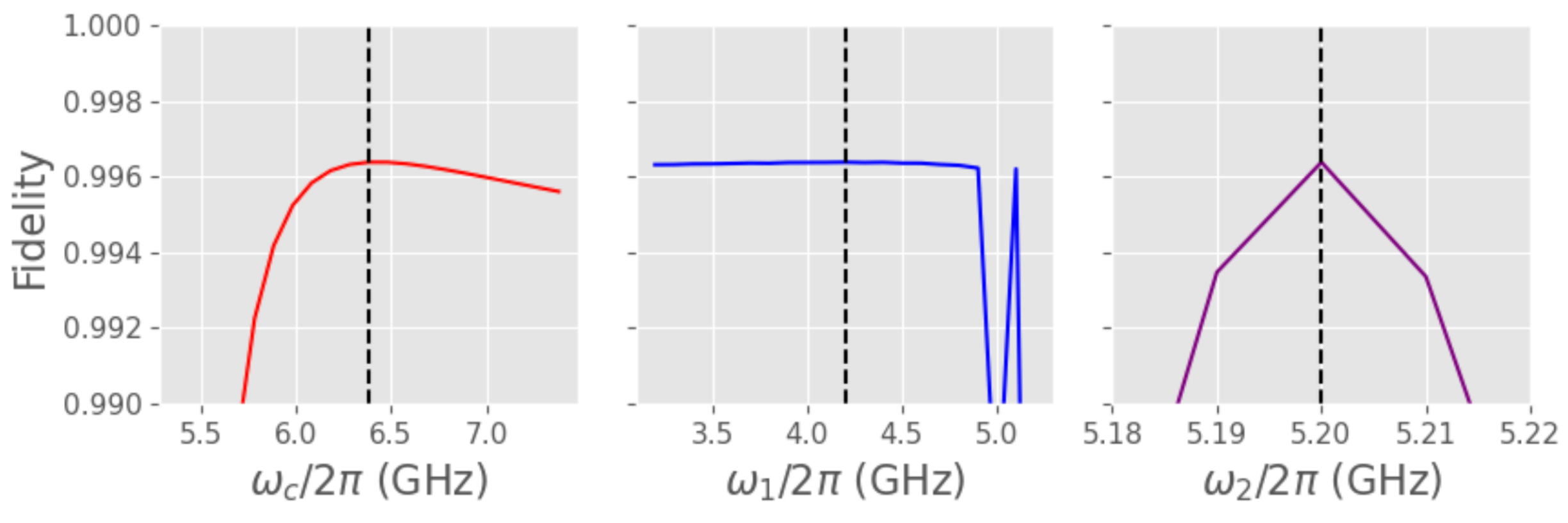}
    \caption{The variation in gate fidelity for a 10 ns gate in response to model fluctuations, in terms of the variation in the \textit{idling} (initial) frequencies for the coupler qubit (red), qubit 1 (blue) and qubit 2 (purple) respectively, when employing the pulses obtained in Fig.~\ref{fig:fig-2} (RL + Auto-diff) and while the other parameters are kept unchanged. The black dotted vertical lines indicate the idling parameter for which the optimized pulses were found.}
    \label{fig:fig-4}
\end{figure}

It is important to note that, in practice, creating pulses with abrupt transitions is challenging and should be substituted with gradual rise and fall transitions. The Appendix E examines the effects of finite rise and fall times for each step using a specific example as well as variations in step size, showing that the inclusion of such effects does not drastically reduce gate fidelity.

Finally, we discuss the prospects of the proposed method for experimental implementations. In this context, we investigate the applicability of the pulse shapes to devices where the transition frequencies of the qubits do not exactly match the parameters of the assumed model, c.f.~Eq.~\ref{eqn:hamiltonian}. 
In Fig.~\ref{fig:fig-4}, we show the robustness of the optimization against frequency fluctuations at the idle point. We consider the optimized pulse that the RL + Auto-diff method found for a set of initial qubit frequencies (shown with black lines), at which the training was done. Then we apply the trained and optimized pulses to a circuit with variations in the idle point qubit frequency, shown along the x-axes. During the gate, the coupler frequency is applied according to the optimized result.
The plots show that the fidelity is maintained up to variations of $\sim10\%$ for $\omega_c$ and $\omega_1$, while it is more sensitive to $\omega_2$. Although this demonstrates the gate's response to parameter uncertainties, the trained RL model can also be re-trained to adjust for such parameter variations. For example, the gate fidelity achieved with the optimized pulse for the coupler's initial frequency of $6.38$ MHz remains robust even when the initial frequency drifts from $6.38$ MHz to $6.10$ MHz, which corresponds to a transverse coupling rate of approximately $0.5$ MHz as shown in Fig.~\ref{fig:fig-supp-1} (a similar transverse coupling strength of 0.3 MHz was considered in~\cite{Barends2014Apr}). This shows that while our optimization scheme is applicable to such bias points while maintaining high fidelity, it can further be improved by retraining at those bias points.
In practical scenarios where the gate pulses developed are utilized in actual experiments, it is anticipated that significant model bias will be observed. These biases can be corrected to accommodate minor parameter deviations encountered in practical applications. However, it is important to recognize that if the system parameter drift is extreme, the resultant pulses may vary substantially. For example, it has been noted that when the RL agent limits the control parameter space to $\omega_c(t)/2\pi = [5.2, 6.38]~ \text{GHz}$ instead of $\omega_c(t)/2\pi = [4.2, 6.38]~ \text{GHz}$ (discussed above), the resulting gate pulses exhibit distinct characteristics with markedly different leakage behavior [see Appendix D].

\section{Conclusion}
In summary, we illustrate an RL-driven methodology for the design of rapid and nonintuitive pulse sequences to execute a two-qubit CZ-gate within a tunable coupler architecture. We show that, grounded solely on penalty or reward considerations, the artificial agent can assimilate effective strategies and unveil realistic parameter configurations for the modulation of coupler piecewise constant flux pulses. We also combine RL with gradient-based optimal control that results in an improved optimizer culminating in ultrafast CZ-gate with high fidelity with error much below the surface code error threshold while maintaining a very short gate time. This is an improvement in gate duration of $\sim6$ and $3$ times, respectively, compared to the CZ gate implementations with tunable coupler in~\cite{Sung2021Jun} with a 60 ns long CZ gate and 34 ns in the surface code implementation with Google's Sycamore processor~\cite{google2023suppressing}. Incorporating experimental data directly into the training process of the RL agent would obviate the necessity for precise simulation models, thereby facilitating the agent's capacity to adjust to device impairments and temporal parameter fluctuations. In this regard, employing the pulse generated by the RL on the simulator as the starting point and subsequently applying a gradient-based method with experimental data would represent a favorable approach.

\begin{acknowledgments}
This work received support from the German Federal Ministry of Education and Research via the funding program Quantum Technologies - from basic research to the market under Contract No.~13N16182 MUNIQC-SC. It is also part of the Munich Quantum Valley, which is supported by the Bavarian state government, with funds from the Hightech Agenda Bayern Plus. BS thanks Lukas Heunisch for useful discussions.
\end{acknowledgments}

\appendix

\section{BASIC THEORY OF REINFORCEMENT LEARNING}
\noindent \textbf{Reinforcement Learning  Workflow}:  Reinforcement Learning (RL) has emerged as an important domain within Machine Learning (ML), notably marked by groundbreaking advancements from DeepMind and Google~\cite{silver_mastering_2017, silver_mastering_2016}. The unique feature of RL lies in its adaptive and iterative learning process, in which the RL agent adjusts its strategy based on real-time consequences within the dynamic environment. In contrast to supervised and unsupervised machine learning methods, which rely on (precollected) labeled and unlabeled datasets for training, RL derives its knowledge through continuous interaction with the environment and evaluation via a reward function. This makes RL particularly well suited for tasks involving control or decision making compared to supervised and unsupervised learning techniques.

\begin{figure}[!hbt]
    \centering
    \includegraphics[width=0.7\linewidth]{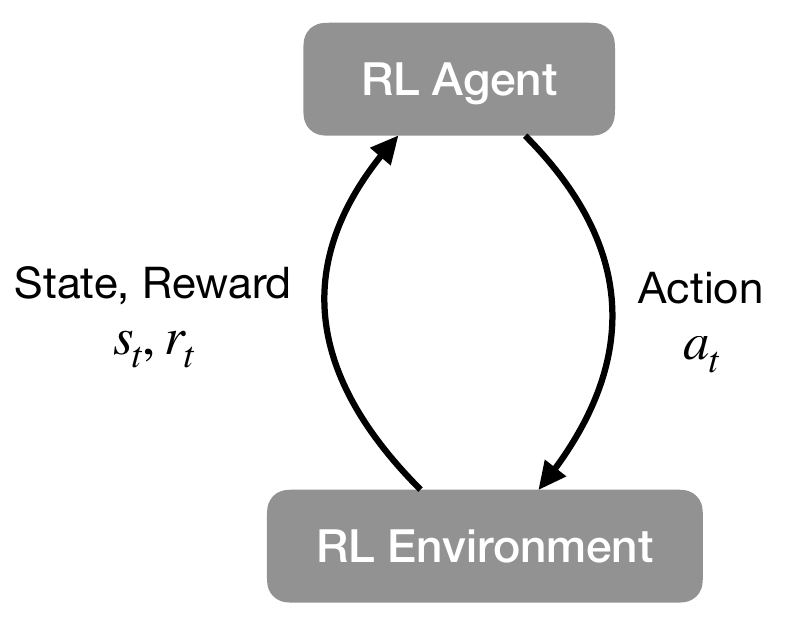}
    \caption{The basic workflow of RL. }
    \label{fig:suppl-rl-workflow}
\end{figure}

The basic RL workflow is shown in Fig.~\ref{fig:suppl-rl-workflow}. It comprises an RL agent that determines what action to take at a specific timestep $t$ in the RL environment (which encodes the system to be controlled), resulting in a modification of the state of the environment. The agent then observes this altered state (which often includes only partial information about the environment), denoted $s_t$. The result or impact of the action, whether beneficial or detrimental, is evaluated by the reward $r_t$ calculated based on the system's new behavior following the action. The RL agent is generally configured as a neural network model, trained through extensive agent-environment interactions to achieve optimal control. The cumulative sum of rewards accumulated over an episode (time length for which we want to learn the control sequences) is used as a metric to adjust the model's weights and biases. The combination of optimized parameters encapsulates the complex rules required to adapt actions based on observed state changes. This set of rules, known as the policy, serves as the decision-making algorithm for the RL agent. \\\\

\noindent \textbf{General Theory}: The policy can be classified into two forms: deterministic and stochastic. 
In the deterministic approach, the action of the RL agent is precisely determined by the policy parameters, denoted as $\theta$, given the state $s_t$ at time $t$. This deterministic relationship is expressed as $a_t = \mu_\theta (s_t)$ (often used for policy evaluation).  Alternatively, the policy can be stochastic, where actions are sampled from a probability distribution conditioned on $s_t$: $a_t \sim \pi_\theta(\cdot|s_t)$. The RL-agent's objective is to iteratively refine and optimize these policy parameters ($\theta$) to maximize the cumulative discounted rewards along a trajectory $\tau = (s_0, a_0, s_1, a_2, ...)$, 
\begin{equation}
    R(\tau) = \sum_{t=0}^T \gamma^t r_t, 
\end{equation}
where the discount factor $\gamma \in (0, 1)$ modulates the significance of future rewards. The optimization process entails maximizing the expected return over discounted rewards, denoted by $J(\pi) = \mathbb E [R(\tau)]$, with the ultimate goal of achieving the optimal policy $\pi^\star = \mathrm{argmax}[J(\pi)]$.  

A related concept is the value function, which is used to predict the expected cumulative discounted future reward and to assess the effectiveness of a given state $s$ or the state-action pair $(s, a)$ in generating a higher net return. The state value $V_{\pi}(s) = \mathbb E[R_t | s_t = s]$ represents the anticipated return when adhering to the policy $\pi$ from state $s$. In contrast, the action value $Q_{\pi}(s, a) = \mathbb E[R_t | s_t = s, a_t = a]$ signifies the expected return when action $a$ is executed in state $s$ followed by policy $\pi$. The Bellman equations~\cite{sutton2018reinforcement} govern the value functions, which can be solved self-consistently. For instance, the action-value function is expressed as:
\begin{align}
Q_{\pi}(s, a) = \mathbb E\left[ r(s, a) + \gamma \cdot \max_{a^\prime} Q_\theta (s^\prime, a^\prime) \right].
\end{align}
Optimizing the policy encompasses various techniques categorized into three main groups: (a) policy-gradient-based, (b) value-based, and (c) actor-critic-based methods. Value-based approaches, such as Q-learning, aim to maximize value functions by solving the Bellman equations. On the contrary, policy gradient methods employ gradient descent algorithms to optimize policy parameters, given by:
\begin{align}
\nabla_\theta J (\pi_\theta) = \mathbb E \sum_{t=0}^{T} \left[\nabla_\theta \log \pi_\theta(a_t | s_t)  R_t \right],
\end{align}
where $\mathbb E$ represents the expectation value over the trajectory $\tau$. This basic approach can be improved by introducing a baseline function, $b(s_t)$, to reduce the variance in gradient estimation, forming the basis of advanced RL actor-critic algorithms. The objective (loss) function for policy gradient methods to optimize is given by:
\begin{equation}
\mathrm {L}^{\rm PG}(\theta) = \hat {\mathbb{E}}_t \left[ \log \pi_\theta(a_t | s_t) A_t\right],
\end{equation}
where $\pi_\theta$ is a stochastic policy, and $\hat{A}_t = Q(s_t, a_t) - V(s_t)$ is an estimator of the advantage function at timestep $t$, considering $R_t$ as an estimate of $Q(a_t, s_t)$. 
An actor-critic algorithm simultaneously learns a policy and a state-value function, using the value function for bootstrapping to reduce variance and accelerate learning~\cite{sutton2018reinforcement}. The critic updates action-value function parameters, and the actor adjusts policy parameters following the critic's guidance.\\\\

\noindent \textbf{Soft Actor-Critic Algorithm}: The Soft Actor-Critic (SAC) algorithm employed in the current study, is a recently developed actor-critic approach in the realm of RL. What sets SAC apart from other actor-critic methods is its distinctive feature of optimizing the policy in an entropy-regularized manner, rendering it inherently stochastic. In SAC, the policy is trained to strike a balance between expected return and entropy. The intentional introduction of entropy serves the purpose of promoting increased exploration and preventing premature convergence of the policy. 

At each time step in RL regularly adjusted for entropy, the agent is compensated according to the entropy of the policy distribution, 
\begin{equation}
\pi^* = \arg \max_{\pi} \underset {\tau \sim \pi}{\mathbb E}
\Big[{ \sum_{t=0}^{\infty} \gamma^t \bigg( R(s_t, a_t, s_{t+1}) + \alpha \mathcal{H}\left(\pi(\cdot s_t)\right) \bigg)}\Big],
\end{equation}
where $\mathcal{H}(P) = \underset {x \sim P}{\mathbb E} [-\log P(x)]$ is the entropy of the probability distribution $P$, and $\alpha > 0$ is the trade-off coefficient. This equation describes the optimal policy $\pi^*$ that maximizes the expected return over time. The return is the sum of the rewards $R(s_t, a_t, s_{t+1})$ at each time step, discounted by a factor $\gamma^t$ to prioritize immediate rewards over future ones. Additionally, this formulation includes an entropy term $\alpha \mathcal{H}\left(\pi(\cdot|s_t)\right)$, where entropy $\mathcal{H}$ encourages exploration by maximizing the uncertainty (randomness) of the policy. The trade-off coefficient $\alpha$ balances the importance of this entropy against the immediate reward.

The value functions in this setting, $V^{\pi}$ and $Q^{\pi}$, are modified accordingly.
\begin{align}
\nonumber
V^{\pi}(s)  = & \underset {\tau \sim \pi} {\mathbb E}\Big[ \sum_{t=0}^{\infty} \gamma^t \big( R(s_t, a_t, s_{t+1}) \\
&+ \alpha \mathcal{H}\left(\pi(\cdot|s_t)\right) \big) | s_0 = s\Big],\\
\nonumber
Q^{\pi}(s,a)  = & \underset {\tau \sim \pi} {\mathbb E}\Big[ \sum_{t=0}^{\infty} \gamma^t R(s_t, a_t, s_{t+1}) \\ 
&+ \alpha \sum_{t=1}^{\infty} \gamma^t \mathcal{H}\left(\pi(\cdot|s_t)\right)|  s_0 = s, a_0 = a\Big].
\end{align}
The state-value function $V^{\pi}(s)$ takes into account the rewards over time and the uncertainty in the policy (via the entropy term). This function essentially tells us how good it is to be in a particular state $s$ under policy $\pi$. The action value function $Q^{\pi}(s,a)$  indicates how good it is to take a particular action $a$ in state $s$. 
The connection between $V^{\pi}$ and $Q^{\pi}$ is shown by the following equation:
\begin{align} V^{\pi}(s) = \mathbb E_{a \sim \pi}\left[Q^{\pi}(s,a) + \alpha \mathcal{H}\left(\pi(\cdot s)\right)\right].
\end{align}
It states that the value of being in state $s$ (under policy $\pi$) can be computed by taking the expected value of the Q-function over all possible actions $a$ the policy might choose, plus the entropy of the policy at that state. This reflects the idea that $V^{\pi}(s)$ considers all potential actions weighted by their probability under $\pi$.

The Bellman equation is a recursive equation used to estimate $Q^{\pi}(s,a)$. It states that the Q-value for taking action $a$ in state $s$ is approximately equal to the immediate reward $r$ plus the discounted value of the next state's Q-value. The subtraction of $\alpha \log \pi(\tilde{a}'|s')$ from this next state Q-value accounts for the entropy of the policy, promoting exploration. The Bellman equation for $Q^{\pi}$ is estimated by:
\begin{align}
  Q^{\pi}(s,a) & \approx r + \gamma\left(Q^{\pi}(s',\tilde{a}') - \alpha \log \pi(\tilde{a}'|s') \right).
\end{align}
In this formulation, the expected values over the next states \(r\) and the states, $s'$ are taken from the replay buffer, while the subsequent actions \(\tilde{a}'\) are sampled from the policy. 
SAC algorithm undergoes the concurrent learning of a policy \(\pi_{\theta}\) and two Q-functions \(Q_{\phi_1}\) and \(Q_{\phi_2}\). The loss functions for each Q-function, enforcing the minimum Q-value between the two Q approximators, are articulated as:
\begin{align}
L(\phi_i, {\mathcal D}) = & \underset{(s,a,r,s',d) \sim {\mathcal D}}{{\mathbb E}}\left[ \Bigg( Q_{\phi_i}(s,a) - y(r,s',d) \Bigg)^2 \right],\\
\nonumber
y(r, s', d) = & r + \gamma \left( \min_{j=1,2} Q_{\phi_{\text{targ},j}}(s', \tilde{a}') - \alpha \log \pi_{\theta}(\tilde{a}'|s') \right), \\
&\tilde{a}' \sim \pi_{\theta}(\cdots').
\end{align}
The loss function $L(\phi_i, {\mathcal{D}})$ is used to train the Q-function $Q_{\phi_i}$. The goal is to minimize the difference (squared error) between the Q-function's current estimate and the target value $y(r,s',d)$. The target value $y(r, s', d)$ represents the estimated return used to update the Q-function. It combines the immediate reward $r$ with the discounted future return from the next state $s'$, considering the minimum value of the two Q-functions $Q_{\phi_1}$ and $Q_{\phi_2}$. The entropy term $-\alpha \log \pi_{\theta}(\tilde{a}'|s')$ encourages exploration by reducing the target value, thus penalizing certainty in the action selection. The data $(s,a,r,s',d)$ are sampled from a replay buffer $\mathcal{D}$, which stores past experiences to break the correlation between consecutive samples during training. Our SAC agent adheres to the implementation as described in~\cite{sac_paper, stable_baselines3, spinningup}.
    

\section{BIAS POINTS}
\begin{figure}[!hbt]
    \centering
    \includegraphics[width=0.75\linewidth]{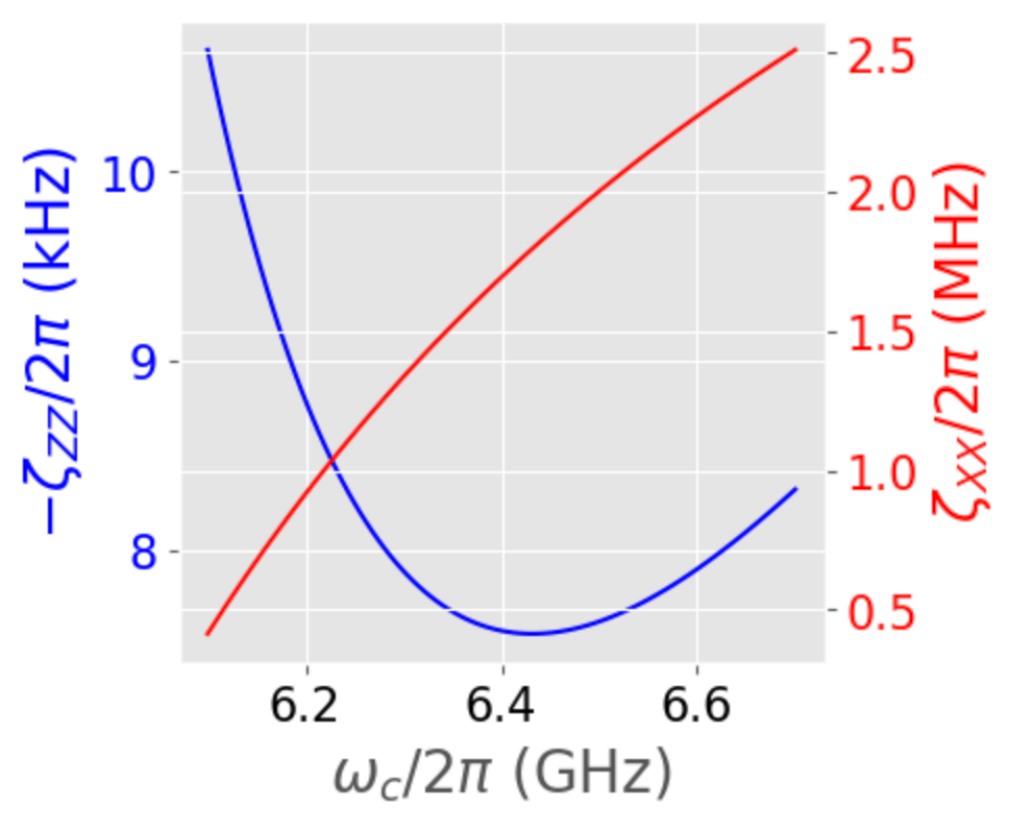}
    \caption{The effective static transverse ($\zeta_{XX}$) and longitudinal ($\zeta_{ZZ}$) couplings. The circuit parameters are stated in the main text.} 
    \label{fig:fig-supp-1}
\end{figure}

\begin{figure}[b]
    \centering  \includegraphics[width=0.9\linewidth]{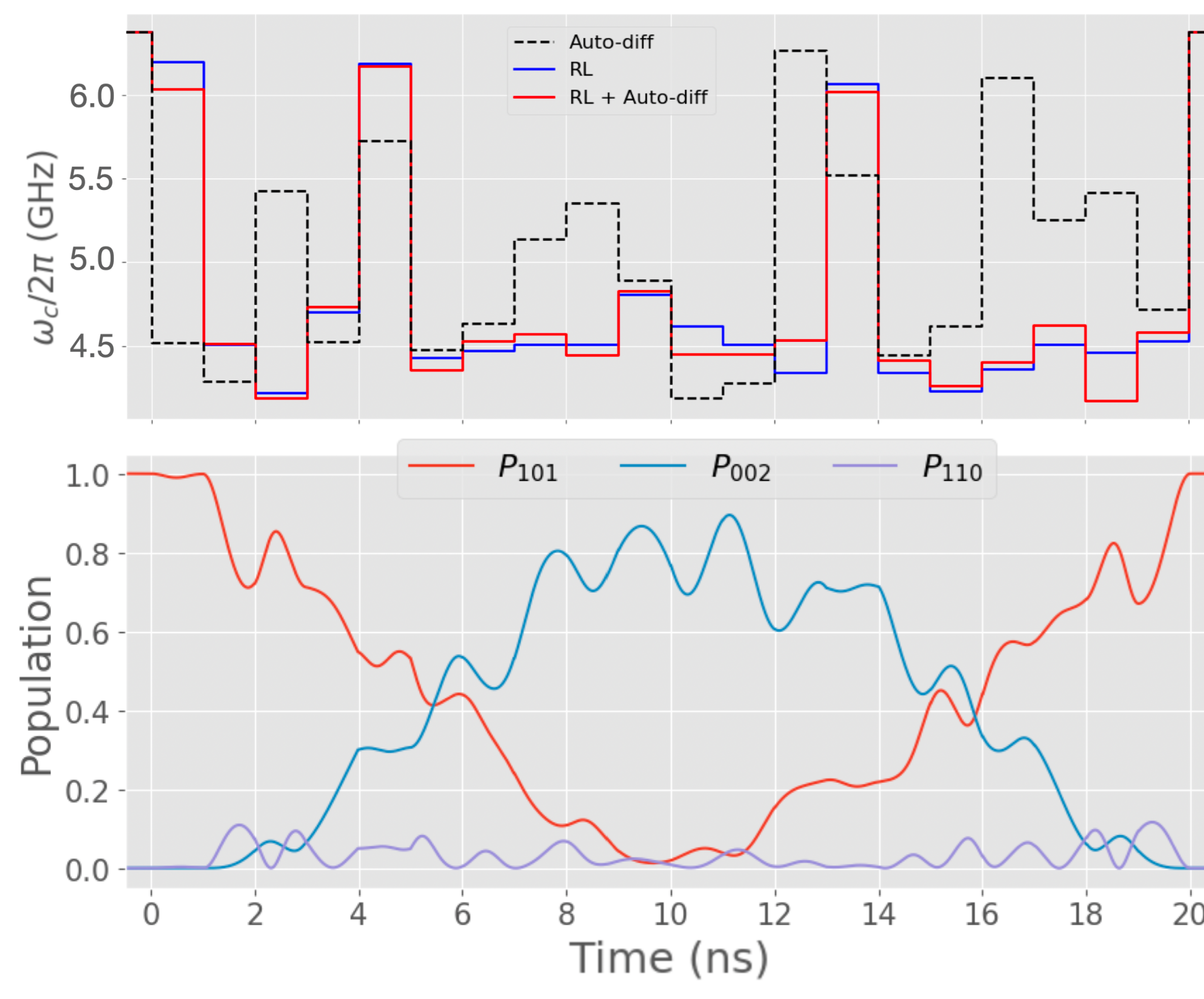}
    \caption{{(Top panel) Control pulses in terms of the coupler frequency derived from the three optimization processes viz.~Auto-diff (black dashed), RL (blue solid) and Auto-diff on RL-optimized pulses (red solid), for a CZ-gate of 20 ns gate time. The parameter search space is restricted to $2\pi\times(4.2 - 6.38)\ \mathrm{GHz}$. Gate time steps are of 1 ns duration. The circuit parameters are given in the main text. (Bottom panel) The population dynamics in the states $|101\rangle$ (red), $|002\rangle$ (blue) and $|110\rangle$ (purple) during the gate, optimized with RL + Auto-diff method. Other leakage states are least occupied.}}
    \label{fig:supp-omegac-4point2}
\end{figure}

\noindent We aim to design the controlled-Z (CZ) gate utilizing the transverse qubit-qubit coupling to induce a phase of $e^{i\pi}$ in the computational state $|101\rangle$ by using nonadiabatic transitions to the non-computational eigenstate $|002\rangle$ and back. Applying a Schrieffer-Wolff transformation, the effective coupler-induced transverse interaction between the data qubits in the dispersive regime is found as,
$\zeta_{XX} = g_{12} + {g_{1c}g_{2c}}\left( {\Delta_{1c}^{-1}} + {\Delta_{2c}^{-1}}\right)/2$,
where $\Delta_{ij} := \omega_i - \omega_j$ denotes the qubit detunings.  
One can see that with proper choice of circuit parameters, the effective two-qubit transverse coupling can be tuned. 
We bias our two-qubit gate circuit in a parameter regime where this effective coupling is small, which is essential for efficient parking of the data qubits (in few MHz range). There is also a residual longitudinal (ZZ) interaction because of dispersive shifts in qubit energies caused by the hybridization of the qubit wave functions, given by $\label{eqn:zeta}
\zeta_{ZZ} = E_{101} - E_{100} - E_{001} + E_{000}$,
where $E_{jkl}$ is the energy eigenvalue of the state ${|jkl\rangle}$. Such residual coupling works as crosstalk in the proper implementation of the gate and should be negligible at the parking point of the circuit. We bias the qubits at the frequencies of $\omega_1/2\pi = 4.2\,\mathrm{GHz}$, $\omega_2/2\pi = 5.2\,\mathrm{GHz}$ and $\omega_c/2\pi = 6.38\,\mathrm{GHz}$. This results in a negligible ZZ-crosstalk ($\zeta_{ZZ}/2\pi = -7.59\,\mathrm{kHz}$) (as shown in Fig.~\ref{fig:fig-supp-1}). As demonstrated in the main text, the optimization remains robust to variations in the coupler frequency at the bias point that corresponds to reduced transverse coupling in the sub-MHz range as well~\cite{Barends2014Apr}.

\begin{figure}[t]
    \centering  \includegraphics[width=0.9\linewidth]{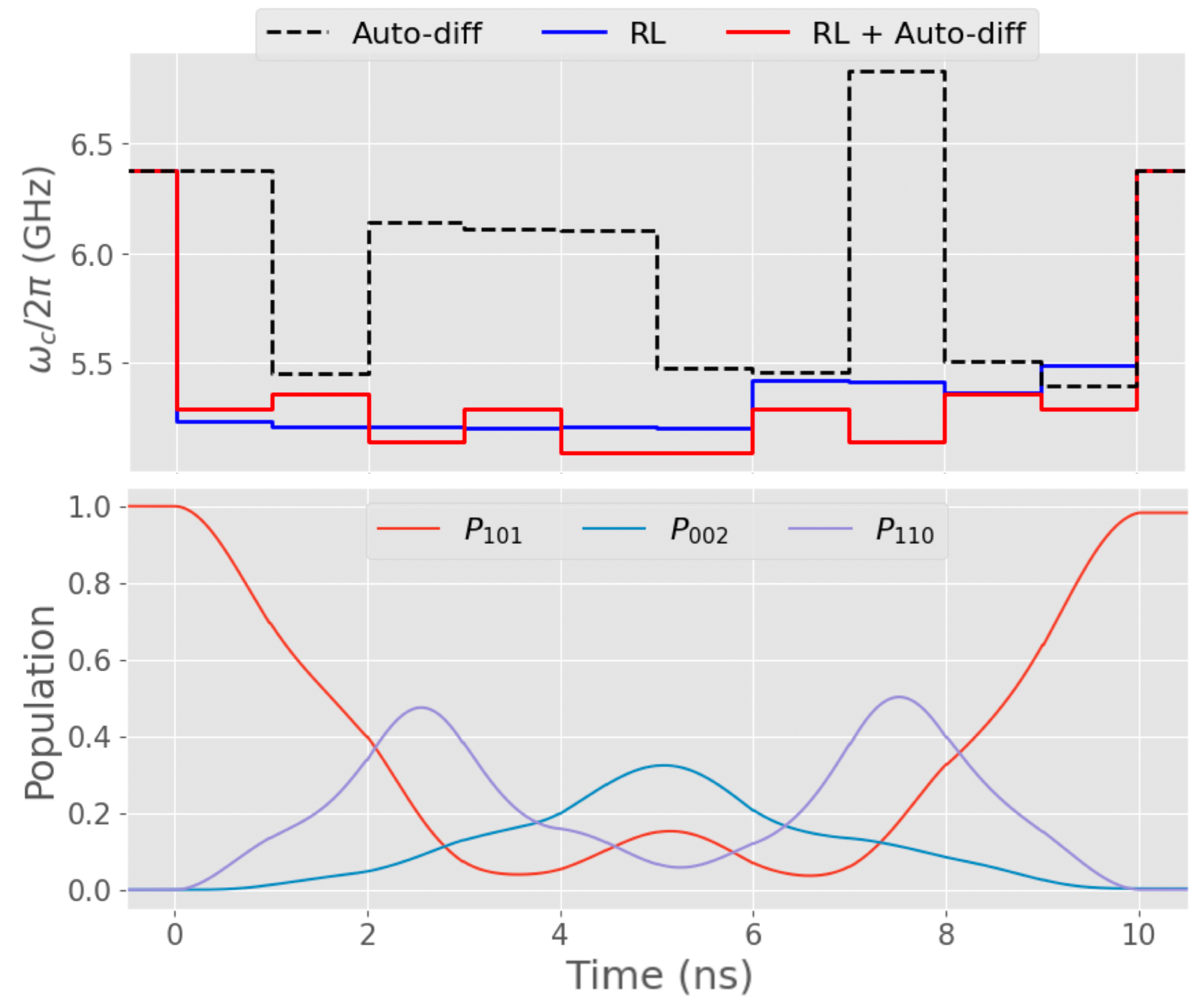}
    \caption{The control pulses (top panel) and the population dynamics (bottom panel) for a shorter pulse of 10 ns gate time for the limited parameter search space of $\omega_c/2\pi = [5.2, 6.38]\ \mathrm{GHz}$. The other circuit parameters are considered as specified in the main text. The control pulses are shown for the three optimization methods, viz.~Auto-diff (black dashed), RL (blue solid) and Auto-diff combined with RL-optimized pulses (red solid). The populations of the states $|101\rangle$ (red), $|002\rangle$ (blue) and $|110\rangle$ (purple) are shown.}
    \label{fig:omegac-5point2}
\end{figure}

\section{CZ GATE WITH GATE TIME OF 20 ns}
\noindent In Fig.~\ref{fig:supp-omegac-4point2} we show a comparison of the gate optimization for the case of the $20$ ns gate found by the three different methods viz.~(i) Auto-diff, (ii) RL and (iii) RL + Auto-diff, and for this we specify the limits of the control parameter, i.e.~the coupler frequency to be in the range of $6.38$ ns (the idle point at the dispersive limit) to $4.2$ ns (near the lower qubit frequency). The resultant population dynamics due to the RL + Auto-diff method is shown in the bottom panel. The leakage population is seen to be significantly low throughout the gate operation and further lowered at the end of the gate.


\section{EFFECT OF LIMITED CONTROL PARAMETER SEARCH-SPACE}

\begin{figure}[t]
    \centering  \includegraphics[width=0.9\linewidth]{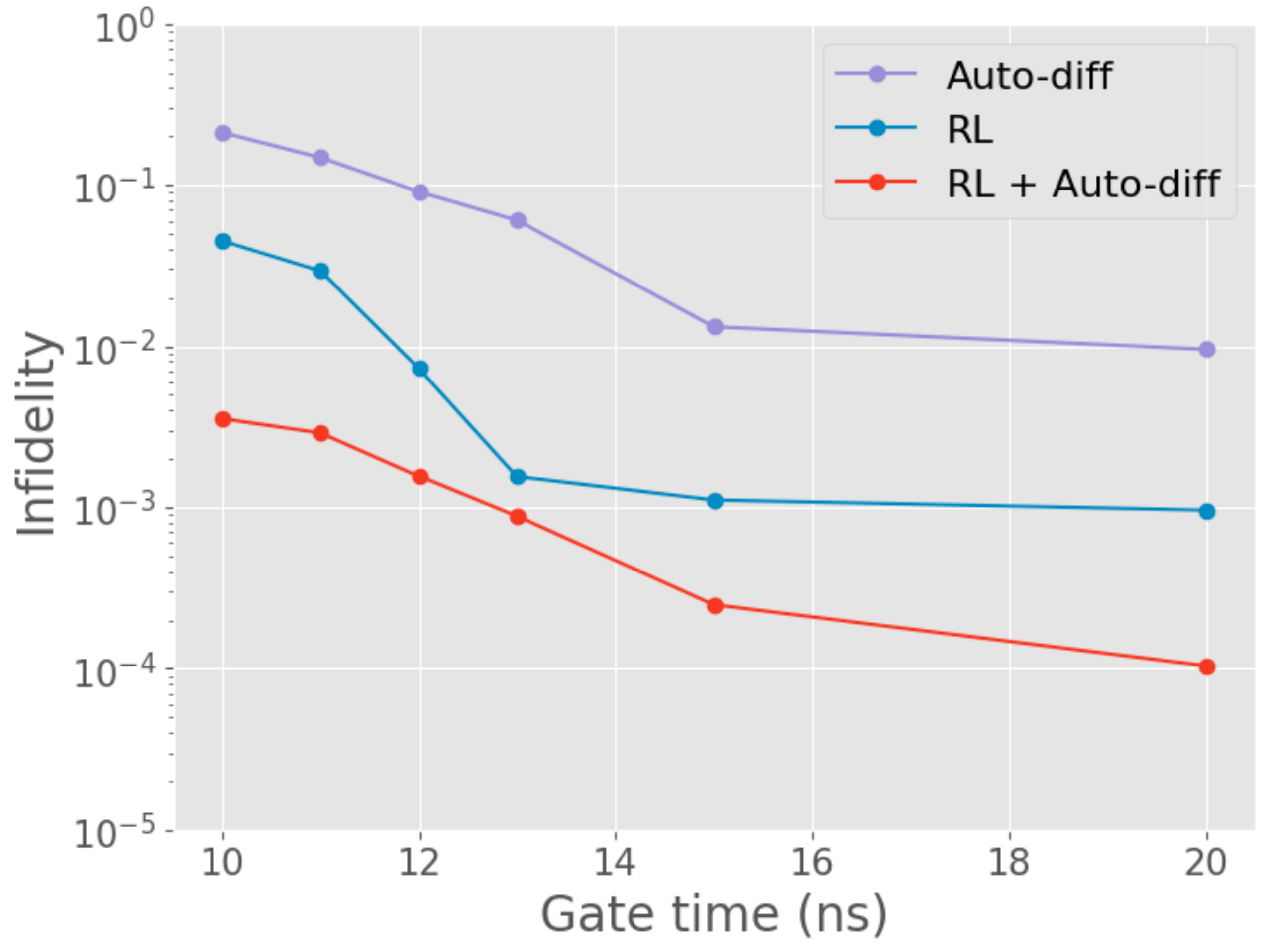}
    \caption{Comparison of the three optimization methods, viz.~Auto-diff (black dashed), RL (blue solid) and Auto-diff on RL-optimized pulses (red solid), shown in terms of gate infidelity with respect to variation in gate time for the limited parameter search space of $\omega_c/2\pi = [5.2,\ 6.38]\ \mathrm{GHz}$. }
    \label{fig:supp-Comparison-omegac-5point2}
\end{figure}

\noindent It has been observed that the control pulses and the consequent gate dynamics vary notably based on the extent of the parameter space accessible to the RL agent. The pulses and dynamics presented in the main text as well as the ones in Fig.~\ref{fig:supp-omegac-4point2} utilize the parameter space where $\omega_c/2\pi \in [4.2, 6.38] \ \text{GHz}$. In contrast, Fig.~\ref{fig:omegac-5point2} shows the pulses and population dynamics for the 10 ns CZ gate found using the Auto-diff, RL and RL + Auto-diff methods with a restricted parameter search space of $\omega_c/2\pi \in [5.2, 6.38] \ \text{GHz}$, where the coupler frequency ranges from the dispersive bias point to the higher data qubit frequency. Compared to this, Fig.~\ref{fig:fig-2} in the main text demonstrates a significant reduction in the leakage population in the coupler during the gate operation while the controls are allowed to explore larger parameter space. In Fig.~\ref{fig:supp-Comparison-omegac-5point2}, we present a comparison of infidelities resulting from the optimizations using the discussed methods, in relation to the change in gate time for the limited parameter space $\omega_c/ 2\pi \in [5.2, 6.38] \ \text{GHZ}$; this can be contrasted with Fig.~\ref{fig:fig-3} from the main text.

\begin{figure}[b]
    \centering   \includegraphics[width=0.9\linewidth]{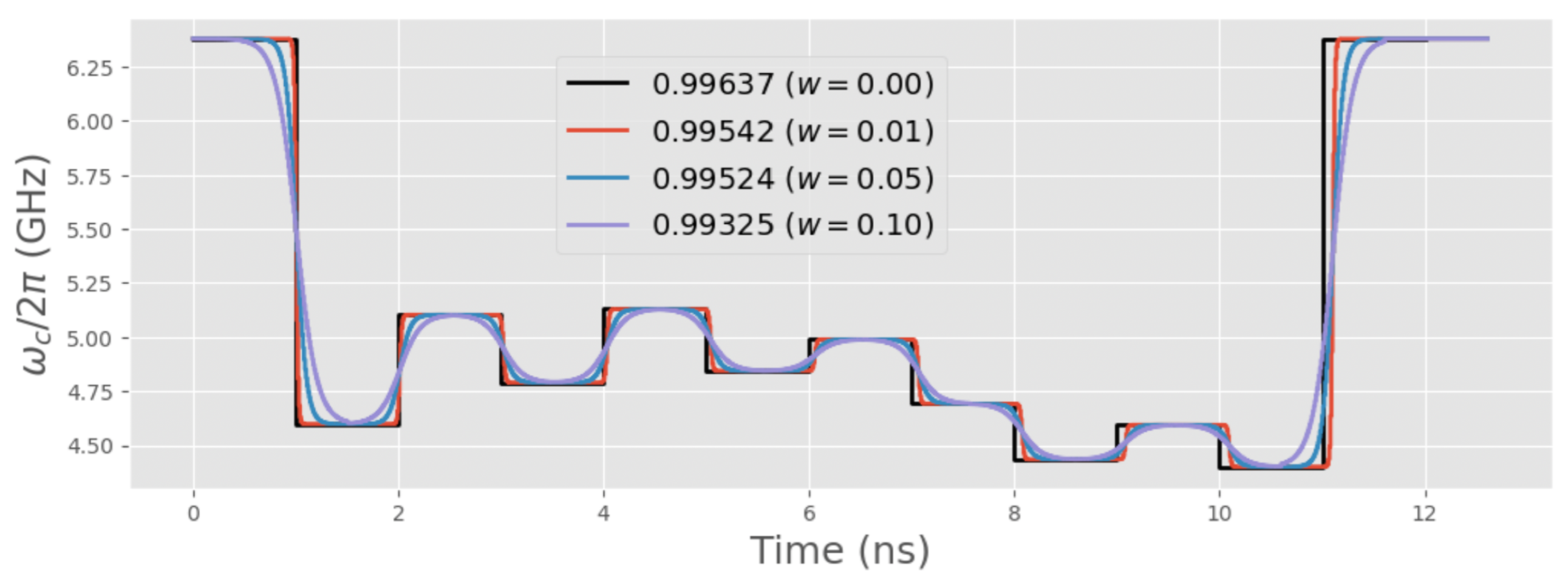}
    \caption{The influence of finite rise and fall times on the gate fidelity is illustrated as a function of w in Eq.~\ref{eq:supp-rise-and-fall} for a 10 ns gate, where higher values of w indicate smoother edges. The inset shows the fidelities corresponding to the values of w. Other parameters are the same as considered in the main text.}
    \label{fig:supp-rise-and-fall}
\end{figure}

\begin{figure}[!t]
    \centering   \includegraphics[width=0.9\linewidth]{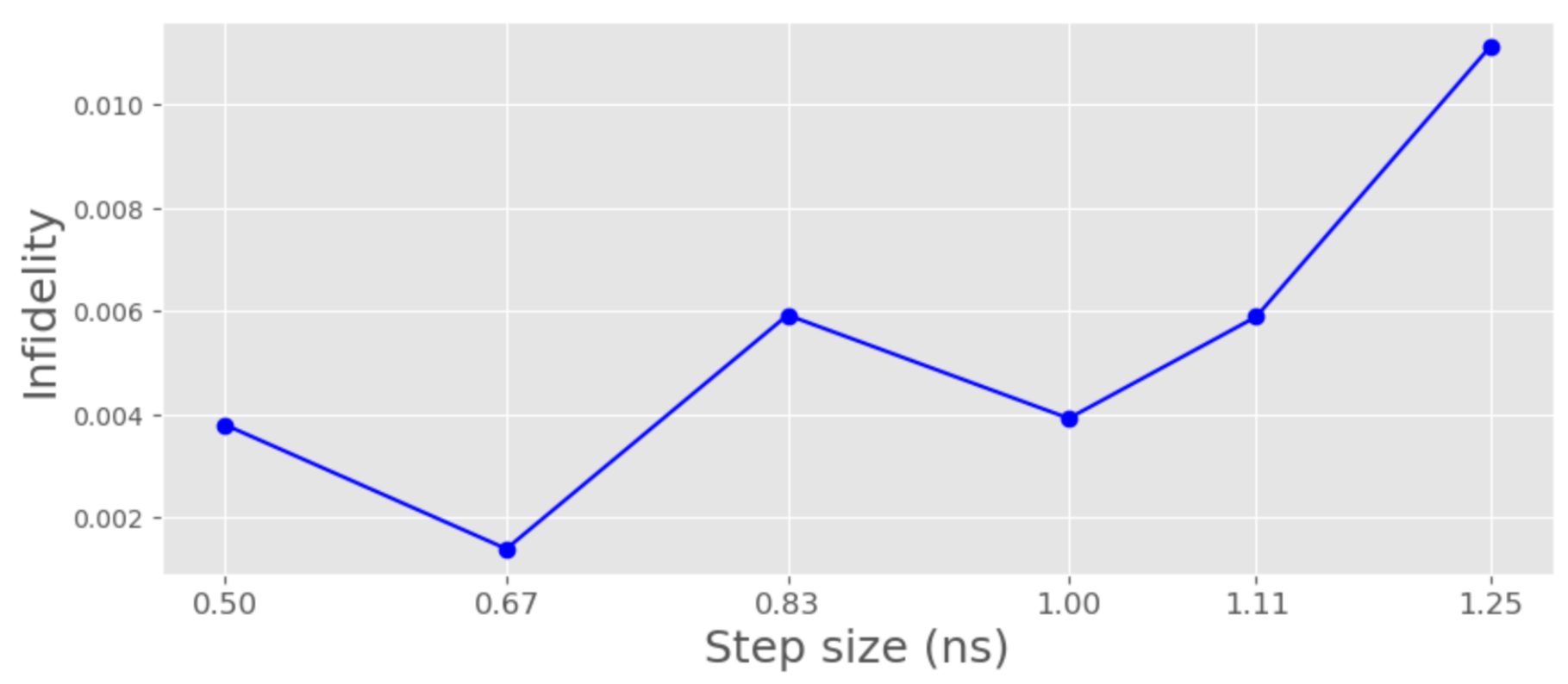}
    \caption{The effect of step size of the control pulse on the gate fidelity for the 10 ns gate. Circuit parameters are the same as considered in the main text.}
    \label{fig:effect-of-dt}
\end{figure}

\section{EFFECT OF FINITE RISE AND FALL TIMES AND STEP SIZE}

\noindent We show the effect of finite rise and fall times for the piecewise constant pulse sequences for a 10 ns gate in Fig.~\ref{fig:supp-rise-and-fall}. It can be modeled as a function of width w, which controls the scaling of the exponent in the logistic function:
\begin{equation}
f(t) = \left[1 + \exp\left(-\frac{t - t_0}{\text{w}}\right)\right]^{-1}. \label{eq:supp-rise-and-fall}
\end{equation}
When w is large, the term $\frac{t - t0}{\text{w}}$ changes slowly, so the exponential term varies slowly, leading to a gradual transition. When w is small, the term $\frac{t - t0}{\text{w}}$ changes rapidly, making the exponential term change quickly, leading to a rapid transition. 
This is a good approximation of practical filtering scenarios similar to also discussed in detail in~\cite{Oh2002Apr}. The corresponding gate-fidelity shows that for a fractional rise and fall time, the fidelity is not significantly altered. While~\cite{Oh2002Apr} reported a quadratic growth of error in rise and fall times, our scheme appears to show a better scaling.

We also show the effect of the step size of the piecewise controls on the optimization with RL+Auto-diff method for a 10 ns long pulse in Fig.~\ref{fig:effect-of-dt}, which shows an overall increasing behavior of gate error. 

\begin{figure}[!hbt]
    \centering
    \includegraphics[width=0.9\linewidth]{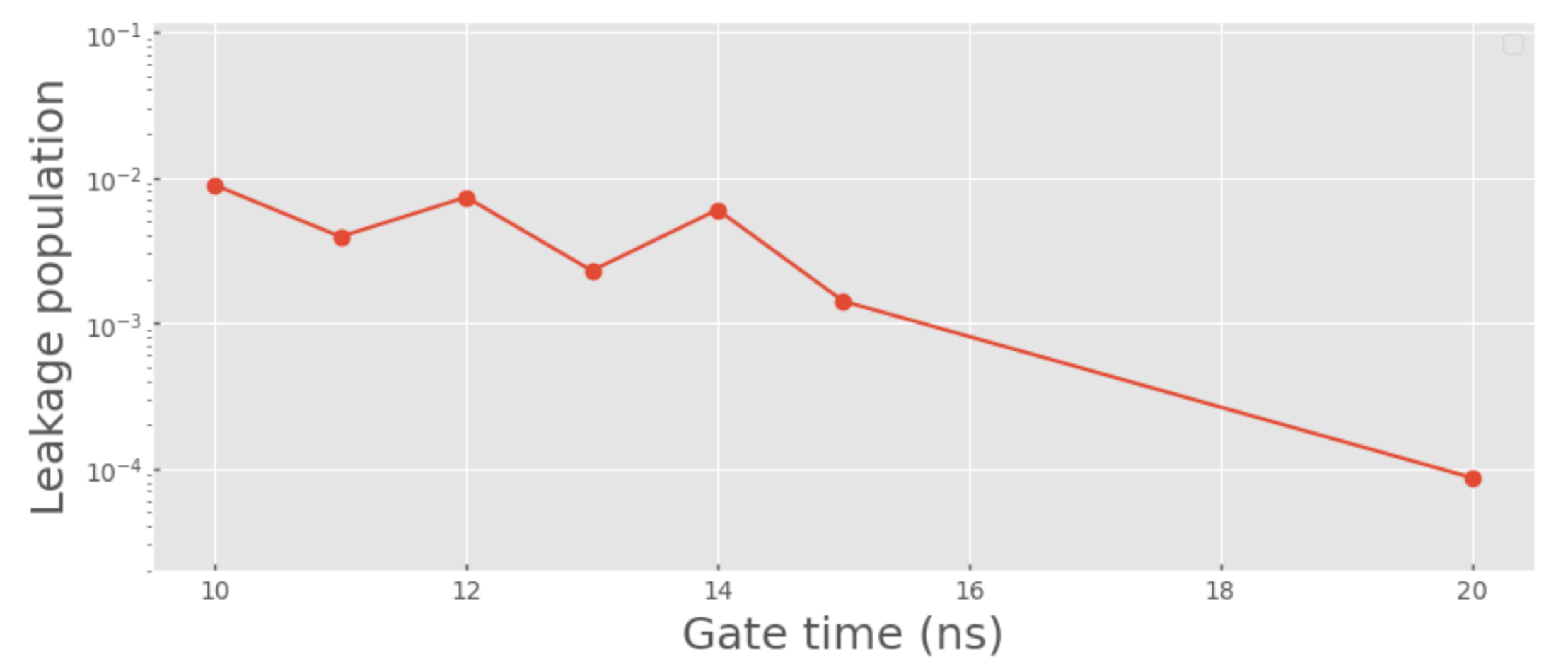}
    \caption{Leakage population outside the computational subspace for the RL+Auto-diff optimized pulses plotted with respect to gate duration. Circuit parameters are the same as stated in the main text.}
    \label{fig:supp-leakage}
\end{figure}

\section{LEAKAGE POPULATION}
\noindent Fig.~\ref{fig:supp-leakage} shows the leakage population outside the computational subspace at the end of the gate as a function of gate time for the RL+Auto-diff optimization, demonstrating an overall decreasing trend with a value of $8\times10^{-5}$ at 20 ns.

\bibliography{reference}

\end{document}